\documentclass[a4paper,11pt]{article}
\usepackage{jinstpub} 
\usepackage{lineno}

\title{\boldmath Contribution of gaseous tritium source rear wall to tritium $\beta-$spectrum in Troitsk nu-mass experiment}

\author{V.\,G.~Chernov}
\author[1]{and V.\,S.~Pantuev\note{Corresponding author.}}
\affiliation{ Institute for Nuclear Research RAS,\\117312 Prospekt 60-letiya Oktyabrya 7a, Moscow, Russia}

\emailAdd{pantuev@inr.ru}

\abstract{We evaluate the contribution of electron scattering on the rear wall of a windowless gaseous tritium source in the Troitsk nu-mass setup. There is a finite probability for them to scatter, return back and reach the detector. A such additional component distorts the measured electron spectrum. We calculate the scattering on the rear side of the vacuum pipe by the GEANT4 simulation and trace the electrons back through the magnetic fields to the spectrometer and  registration system. The contribution is a few orders of magnitude smaller compared to the  original spectrum at the end of the $\beta$-spectrum of 18.6 keV but reaches 2\% at 11 keV. Taking into account this rear wall scattering we can fit well the measured spectrum in search for a sterile neutrino component in the tritium beta decay.}

\keywords{Large detector systems for particle and astroparticle physics}

\begin{document}
\maketitle
\flushbottom

\section{Introduction}
\label{sec:intro}
The goal of the Troitsk nu-mass experiment is a search for sterile neutrinos in the tritium beta decay, ref.~\cite{Abdurashitov:2015jha}. One of the main parts of the setup is the windowless gaseous tritium source (WGTS)  about 5 meters long. A set of solenoids in WGTS forms a magnetic bottle . Electrons at small pitch angle to the magnet axis will leave the WGTS towards the spectrometer or in the opposite direction. The spot of electrons that escape from the WGTS in the direction opposite to the spectrometer (rear wall) increases in size in the fringe field and then scatters on the vacuum pipe. At some combination of scattered angle and final energy, electrons may return to the WGTS and spectrometer and reach the detector. This contamination modifies the measured spectrum. In addition to  direct penetration to the rear wall, there is also a finite probability for trapped electrons to reach the rear wall after multiple electron scattering in the working gas, reference~\cite{Nozik:2020xvw}.

In this article we estimate contribution of the  the rear wall to the measured integrated tritium spectrum and demonstrate its influence to the spectrum. We also want  to stress that the rear wall problem is crucial for measurements similar to ours, such as the TRISTAN project, which is an extension of the KATRIN experiment, reference~\cite{TRISTAN}.
\section{Troitsk nu-mass}
\label{sec:troitsk}
\begin{figure}[htbp]
	\centering
		\includegraphics[width=.95\linewidth]{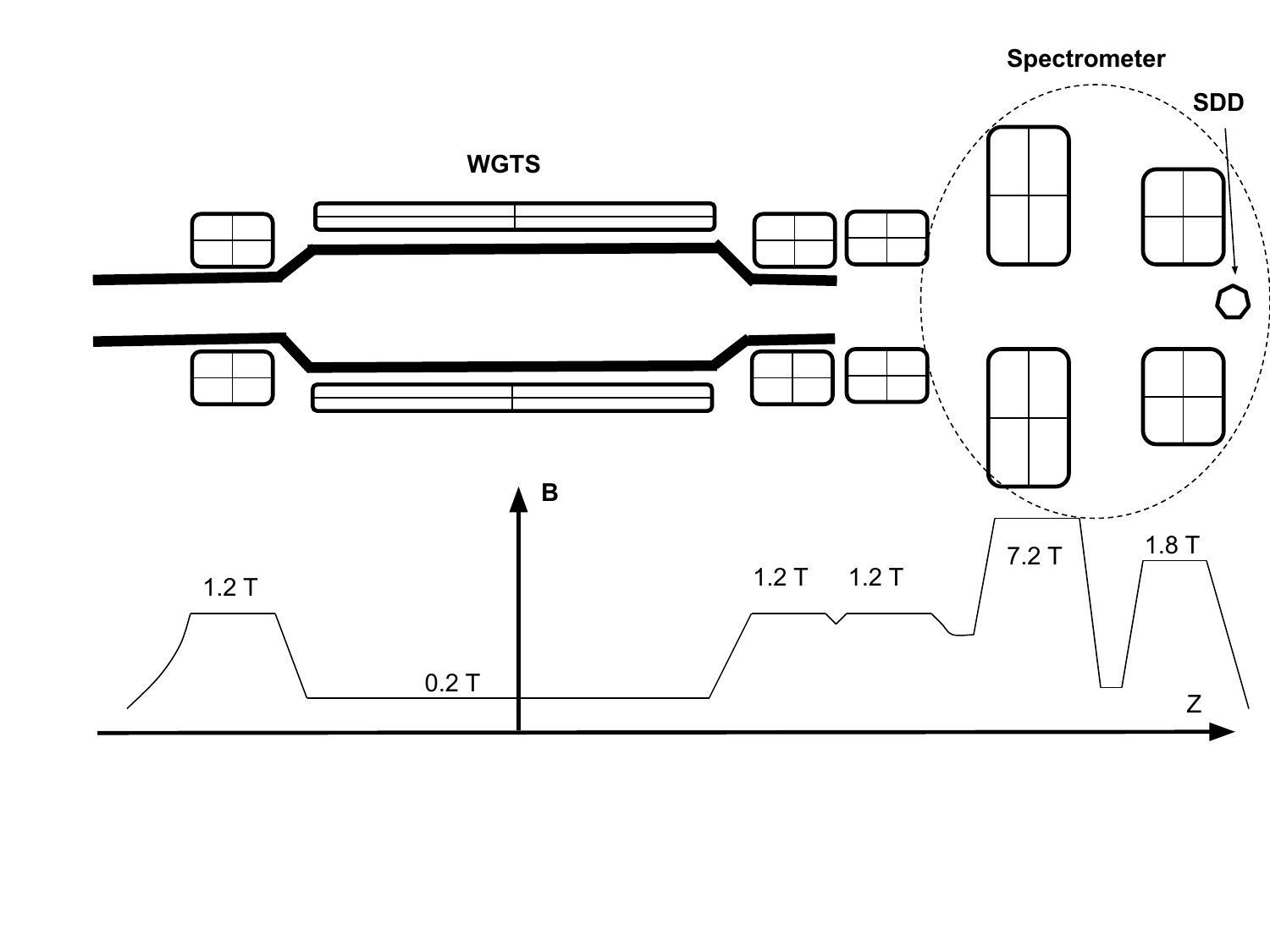}
	\caption{Sketch of the Troitsk nu-mass setup (top), and magnetic fields, B in Tesla, configuration (bottom). The spectrometer is symbolically represented by an entrance pitch magnet at 7.2~T and a detector magnet at 1.8~T.}
	\label{fig:numass}
\end{figure}

The experiment consists of two main components, figure~\ref{fig:numass}, which are a Windowless Gaseous Tritium Source and an Electrostatic Spectrometer with Magnetic Adiabatic Collimation (MAC-E filter), reference~\cite{Abdurashitov:2015jha}. The spectrometer entrance pitch magnet is at 7.2~T. The magnetic field in the detector location is  1.8~T. A gaseous source with freely circulating radioactive gas allows one to avoid solid state effects associated with a "substrate" or "window". 

 The magnetic field in the WGTS has the shape of a magnetic bottle formed by the superconducting solenoids with a field value of 0.2~T in the main central part with a diameter of 50 mm and 1.2~T magnets at the edges with an inner diameter of 20 mm.  To describe the motion of electrons in a magnetic field it is necessary to use a simple formula for transformation of the particle azimuthal angle $\theta$ relative to the field direction with the field strength, $B$.
\begin{equation*}
	 \frac{sin(\theta_1)}{sin(\theta_2)} = \sqrt{B_1/B_2}.
\label{eq:1}	
\end{equation*}
We can estimate the maximum angle of electrons produced in the WGTS at which they will escape from the bottle. In this case $sin(\theta_1)=1$  and $\theta_2= arcsin(\sqrt{B_2/B_1})$. With $B_1$=1.2~T and $B_2$=0.2 we get 
 23 degrees relative to the field axis. All electrons at a larger produced angle will be trapped.

Electrons  from the WGTS are transported to the spectrometer by a system of superconducting solenoids.  A pitch solenoid at 7.2~T at the spectrometer entrance sets an additional angular cut of about 9 degrees for the $\beta$-electrons produced in the WGTS in the direction of the spectrometer. 
The MAC-E filter acts as an electrostatic rejector of all electrons with energy less than a given spectrometer potential. 
Measurement at different values of the retarding potential  gives an integrated  energy spectrum. 
 A 7~pixel silicon drift detector (SDD), figure~\ref{fig:numass}, with a total diameter of about 6 mm is used as an electron detector. This SDD similar to the one used in reference~\cite{Mertens:2020mdv}. 

\section{Rear wall simulation}
\label{sec:simulation}
\begin{figure}[htbp]
	\centering
		\includegraphics[width=0.8\linewidth]{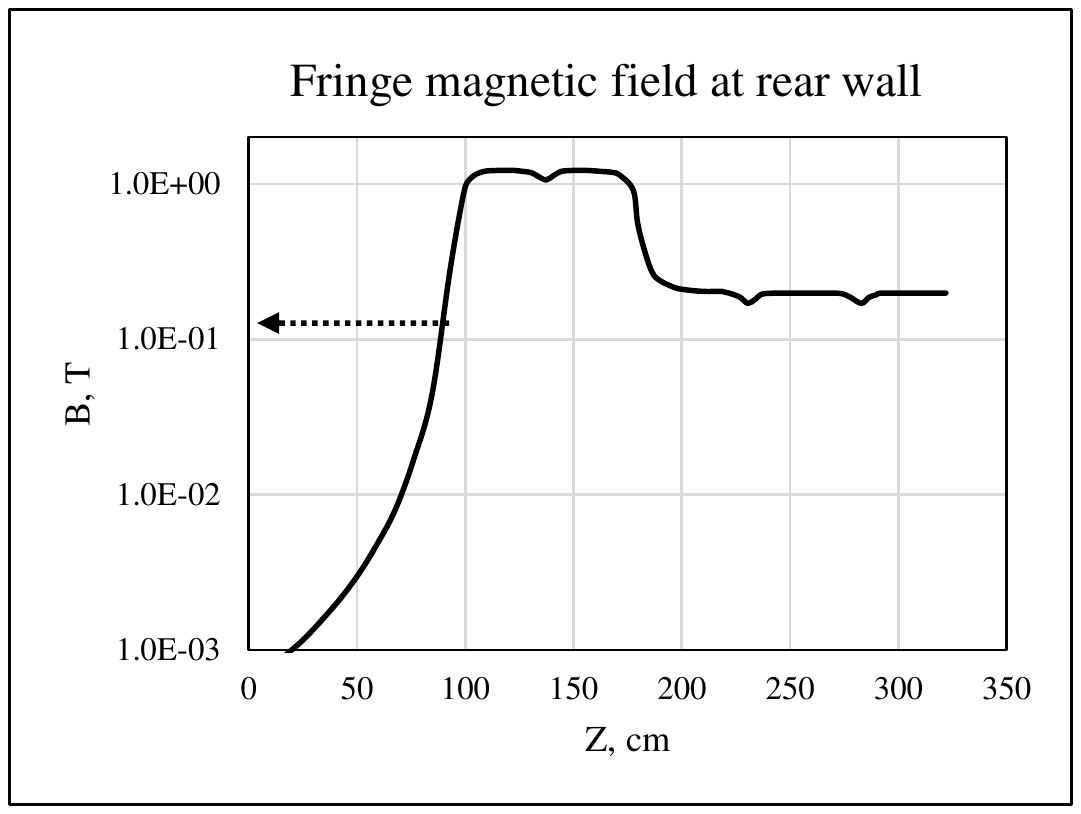}
		\caption{Part of the WGTS magnetic field map with a fringe field  at the rear wall location. The arrow shows the field strength and the $Z$ axis location where scattered electrons may return back to the spectrometer.}
	\label{fig:fringe}
\end{figure}

 The spot of electrons which leave the WGTS  through the last solenoid towards the rear wall 
 increases in diameter in the falling down fringe field, figure~\ref{fig:fringe}. To understand the influence of the rear wall, we start from the opposite direction, from the detector location. The magnetic field flux there at B=1.8~T and a  detector diameter of 6~mm is about 50.9~T$\cdot mm^2$. This corresponds to a spot of  18 mm diameter in the middle of the WGTS with a field of 0.2~T. It also confirms that the wall of the  vacuum pipe with a diameter  of 50 mm is far from this spot. The size of the magnetic flux seen by the detector is enlarged by fringe field at the rear wall and starts touching the 20~mm diameter vacuum pipe at the field of $50.9/(20^2\cdot 3.14/4)$=0.162~T. To escape from the WGTS the electrons have to pass through the rear magnet with a field value of 1.2~T. The maximal azimuthal opening angle for such electrons at  a location with a fringe field of 0.162~T  should be 
$\theta=arcsin(\sqrt(0.162/1.2))=22^{\circ}$. 

\begin{figure}
	\centering
		\includegraphics[width=0.8\linewidth]{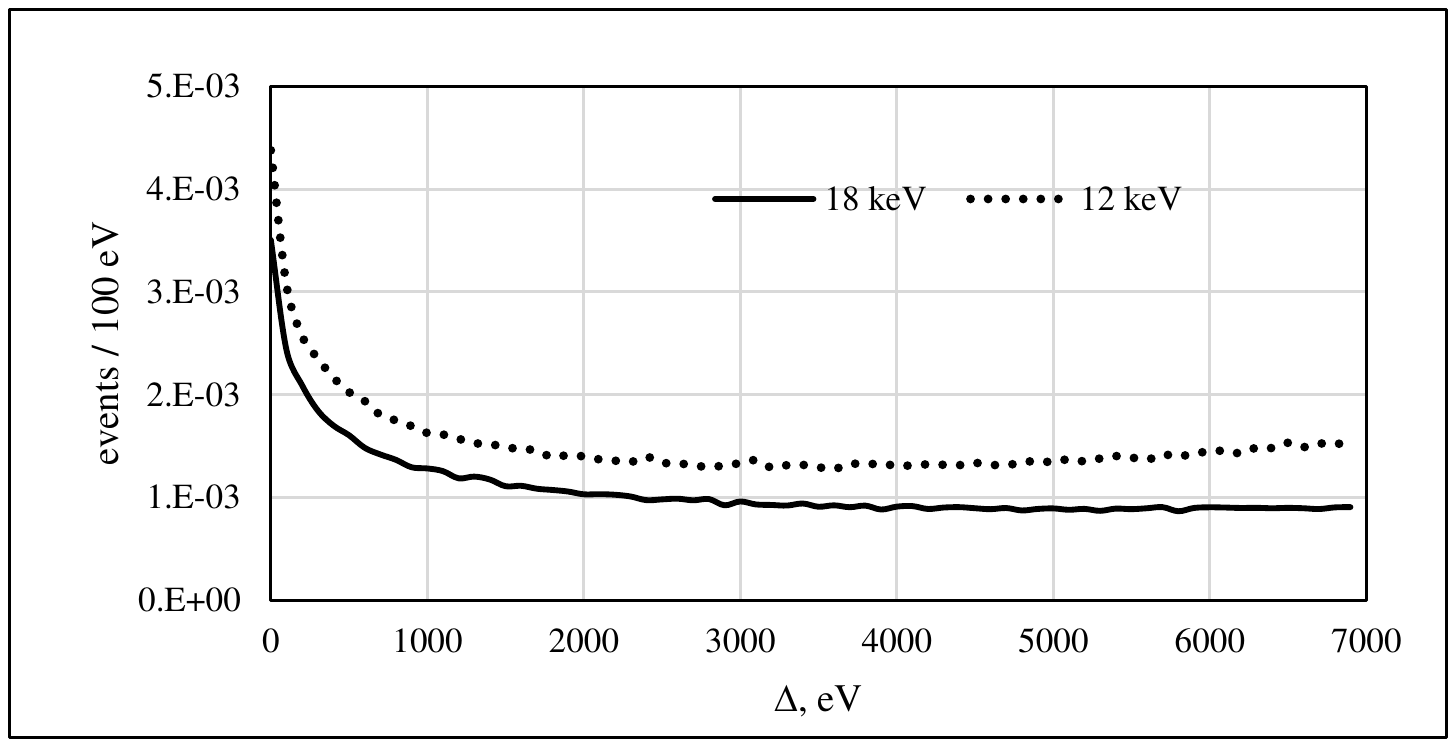}
	\caption{Simulation of electron energy loss spectra at initial energies of 12~keV and 18~keV. Electrons are originally trapped ($\theta > 23^{\circ}$) but then scatter in the direction of the rear wall.  $\Delta$ is the energy loss, the vertical axis shows the number of electrons per 100~eV bin for 4M simulated events.}
	\label{fig:trapping}
\end{figure}

To estimate the influence of the rear wall we did a simple simulation of the scattered electrons on the metal wall 
using GEANT4 framework, ref.~\cite{geant}. We assumed a flat $dN/(cos\theta d\theta)$ angular impact distribution in the range $0-22^{\circ}$. The differential tritium spectrum is used  as an electron energy distribution. We also have to account for the additional component from the electrons that escaped from the WGTS magnetic trap in the rear wall direction. For this purpose, we calculated separately the electron energy loss spectra  after multiple scattering in the working gas, $\Delta=E_{initial}-E_{final}$, similar to the estimation described in reference~\cite{Nozik:2020xvw}. Examples of the calculation for two initial energies are shown  in figure~\ref{fig:trapping}. These spectra were convoluted with the original $\beta$-spectrum. 

Only a small fraction of the scattered electrons can return and reach the detector in the spectrometer.  As mentioned, electrons from the WGTS  are selected by the  spectrometer pitch magnet  with an azimuth's angle of less than about 9 degrees. For electrons scattered backward at the rear wall  this angle is about the same, $8.6^{\circ}$. The latter value sets electrons angular cut for the accepted from the rear wall.  

\section{Results}
\label{sec:results}
\begin{figure}[htbp]
	\centering
		\includegraphics[width=0.8\linewidth]{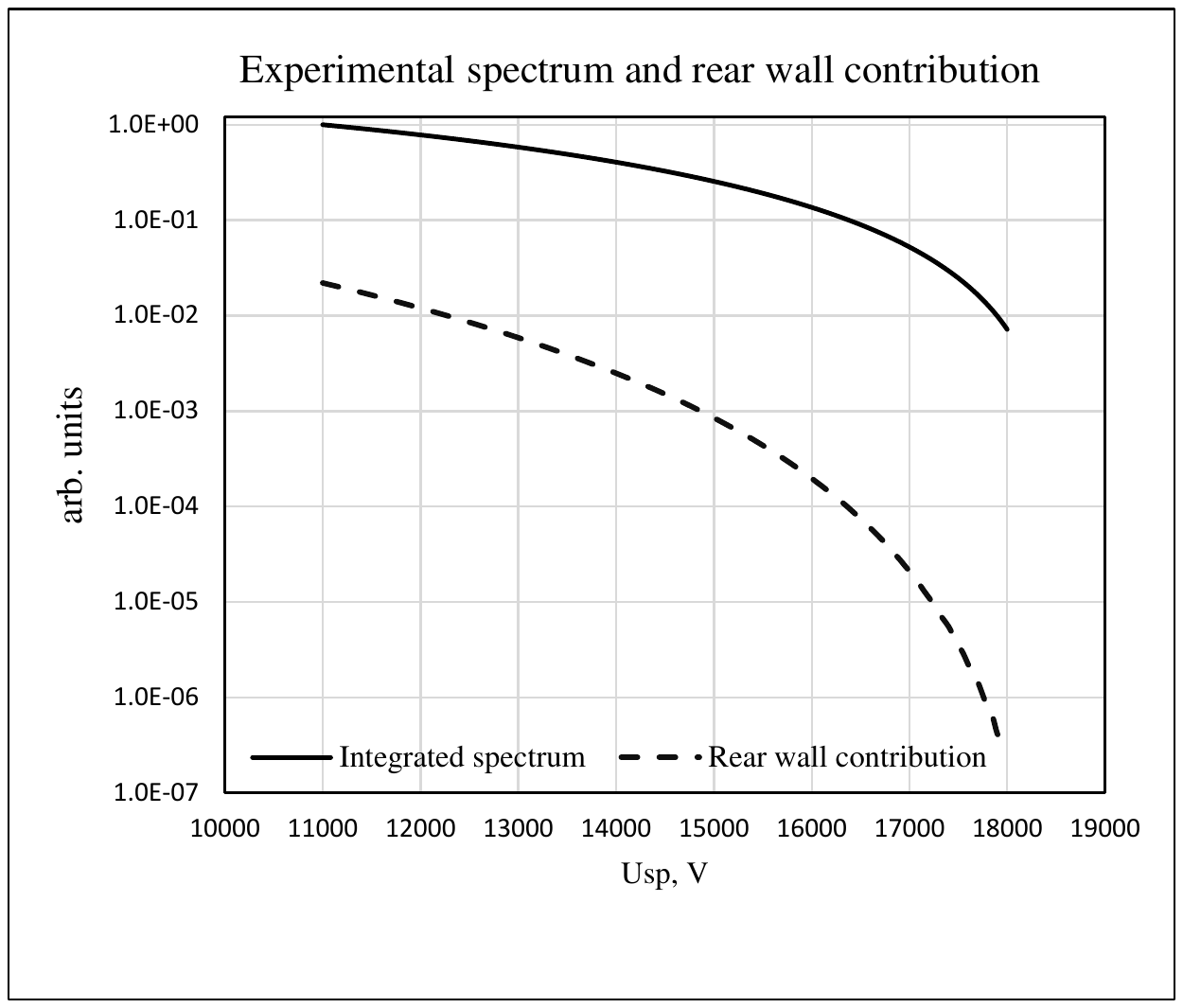}
	\caption{Shape of the measured integrated tritium spectrum and the rear wall contribution depending on the spectrometer potential.}
	\label{fig:rw_contribution}
\end{figure}

The result of our estimation of the rear wall contribution to the experimental spectrum is shown in figure~\ref{fig:rw_contribution}. At each spectrometer potential the rear wall spectrum was integrated above this value. The effect drops quickly as the spectrometer potential increases. This explains why in all previous measurements near the end point of the electron energy spectrum  it was possible to ignore the rear wall contribution. As soon as we go down from 18.6 keV to 11-12 keV of the spectrum low energy range, the effect becomes larger. It reaches about 2\% of the original integral spectrum. 

\begin{figure}
	\centering
		\includegraphics[width=.8\linewidth]{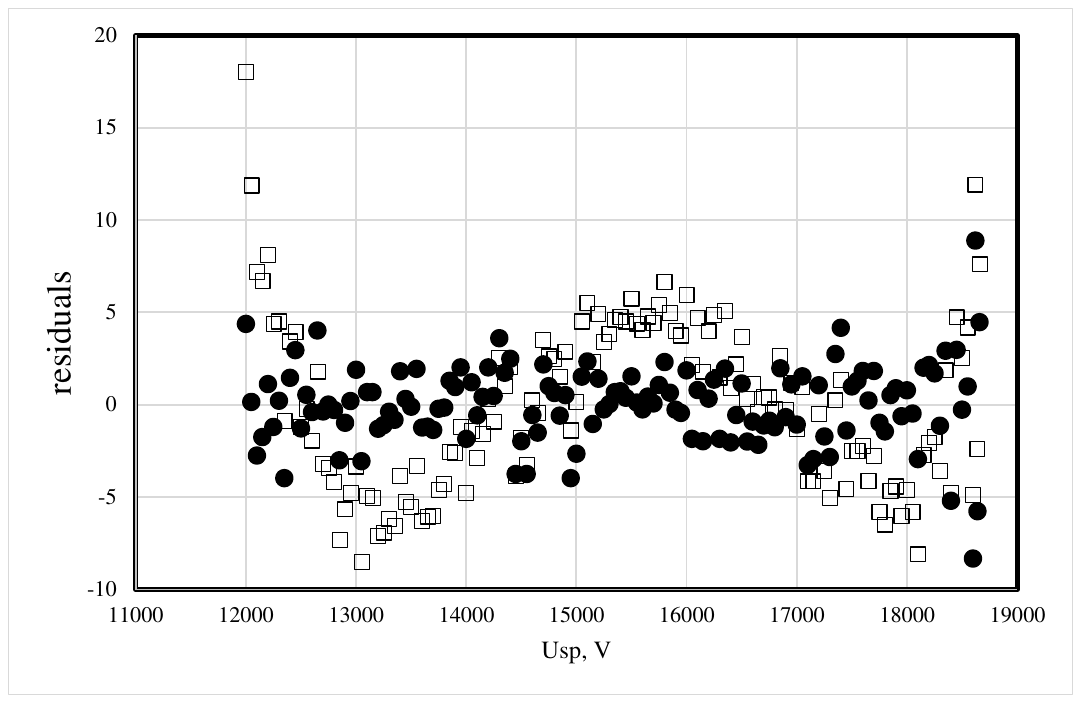}
	\caption{Residuals in sigmas of the fit of experimental data with (black circles) and without (open squares) correction for the rear wall.}
	\label{fig:fit}
\end{figure}

 It is difficult to estimate the absolute value of the rear wall contribution because of uncertainties in geometry, magnetic field configuration, GEANT4 simulation. Thus, we are interested in the shape of an additional spectrum contamination, leaving its normalization as a free parameter.  Figure~\ref{fig:fit} illustrates how the fitting of the experimental spectrum changes without and with the rear wall effect. Without the rear wall correction the fit fails. 

\acknowledgments
We would like to thank Susanne Mertens, who provides us with SDD detector. This work was supported by the Ministry of Science and High Education of the Russian Federation under Contract 075-15-2024-541.


\begin{thebibliography}{99}

\bibitem{Abdurashitov:2015jha}
D.~N.~Abdurashitov {\it et al.},
\emph{The current status of "Troitsk nu-mass" experiment in search for sterile neutrino},
\emph{JINST} {\bf 10} (2015) T10005.
doi:10.1088/1748-0221/10/10/T10005.
arXiv:1504.00544.

\bibitem{Nozik:2020xvw}
A.~Nozik and T.~Hamitov,
\emph{Electron evaporation from magnetic trap in Troitsk nu-mass experiment},
\emph{JINST} {\bf 16} (2021) P05022.
doi:10.1088/1748-0221/10/10/T10005.
arxiv:1234.5678.

\bibitem{TRISTAN}
M Aker et al.,
\emph{KATRIN: status and prospects for the neutrino mass and beyond},
\emph{J. Phys. G: Nucl. Part. Phys.} {\bf 49} (2022) 100501.
doi:10.1088/1361-6471/ac834e
[arXiv:2203.08059]. 

\bibitem{Mertens:2020mdv}
S.~Mertens {\it et al.},
\emph{Characterization of silicon drift detectors with electrons for the TRISTAN project}
\emph{J. Phys. G: Nucl. Part. Phys.} {\bf 48} (2020) 015008.
doi:10.1088/1361-6471/abc2dc
[arXiv:2007.07136].

\bibitem{geant}
https://geant.org/.

\end{thebibliography}
\end{document}